# Chemistry of Silicate Atmospheres of Evaporating Super-Earths

Recommended short title: Silicate Atmospheres of Super-Earths

Laura Schaefer[1,2]

Bruce Fegley, Jr.[1,3]

[1]*Planetary Chemistry Laboratory*

*McDonnell Center for the Space Sciences*

*Department of Earth and Planetary Sciences*

*Washington University in St. Louis*

[2]*laura_s@levee.wustl.edu*

[3]*bfegley@levee.wustl.edu*

**Corresponding Author:**

Bruce Fegley, Jr.

Washington University

Department of Earth and Planetary Sciences

Campus Box 1169

St. Louis, MO 63130-4899

bfegley@levee.wustl.edu

FAX: (314) 935-7361

PHONE: (314) 935-4852

**ABSTRACT:** We model the formation of silicate atmospheres on hot volatile-free super-Earths. Our calculations assume that all volatile elements such as H, C, N, S, and Cl have been lost from the planet. We find that the atmospheres are composed primarily of Na, $O_2$, O, and SiO gas, in order of decreasing abundance. The atmospheric composition may be altered by fractional vaporization, cloud condensation, photoionization, and reaction with any residual volatile elements remaining in the atmosphere. Cloud condensation reduces the abundance of all elements in the atmosphere except Na and K. We speculate that large Na and K clouds such as those observed around Mercury and Io may surround hot super-Earths. These clouds would occult much larger fractions of the parent star than a closely bound atmosphere, and may be observable through currently available methods.
**Keywords:** atmosphere, silicate, exoplanet, chemistry

# 1. INTRODUCTION

There are currently 26 known exoplanets with $M < 20 M_\oplus$. Their compositions are unknown: some may be Earth-like planets (rock + metal), while others may be mini-Neptunes (ice + rock) that have migrated inwards (Barnes et al. 2009). Many of these super-Earths have short orbital periods ($a < 0.5$ AU) and are strongly irradiated by their stars. Tidal forces are also very strong and can generate sufficient heat to melt a silicate planet larger than the Earth (e.g. Jackson et al. 2008a, b; Ganesan et al. 2008). Jackson et al. (2008b) calculated that the heat flux of GJ 876d ($M \sim 6 M_\oplus$) from tidal heating would be $\sim 10^4$-$10^5$ W/m$^2$ and that of Gl 581c ($M \sim 5 M_\oplus$) $\sim 10$ W/m$^2$. For comparison, the heat flux of Io, which is the most volcanically active body in our solar system, is $\sim 3$ W/m$^2$ (Spencer & Schneider 1996).

To date, detections of exoplanet atmospheres have been limited to transiting gas giant planets: e.g., Na, H, C and O in HD 209458b (Charbonneau et al. 2002, Vidal-Madjar et al. 2003, 2004), and Na in HD 189773b (Redfield et al. 2008). However, advances in detection techniques are being made, and detections of the atmospheres of super-Earths may soon be possible. Predictions of spectroscopically active gases are thus timely.

So far, atmospheric models of super-Earths have focused on traditional compositions containing H, C, and N (e.g., Miller-Ricci et al. 2009, Elkins-Tanton & Seager, 2008, Ehrenreich et al. 2006). Given the extreme heating experienced by many super-Earths, we consider here silicate atmospheres produced by vaporization of volatile–free (e.g., H, C, N, S) super-Earths. Although we do not directly model the loss of volatiles, we justify this model in two ways: (1) with literature on evaporation of giant exoplanets and (2) with examples from our own solar system (Io and Venus).

Hot extrasolar giant planets (EGPs) are gas giants with very short orbital periods. One hot EGP (HD209458b) is known to be evaporating (Baraffe et al. 2004, Hubbard et al. 2007). Models of hot exoplanets (e.g., Lecavelier des Etangs et al. 2004, Baraffe et al. 2005, Lecavelier des Etangs 2007) suggest that a hot EGP may lose a significant fraction of its mass over the planet's lifetime. Baraffe et al. (2005) suggest that so-called hot-Neptunes may have begun as Jupiter-sized objects, which have subsequently lost a significant portion of their mass. Lecavelier des Etangs et al. (2004) and Lecavelier des Etangs (2007) suggest that planets at orbits less than 0.03-0.04 AU may be quickly and completely vaporized unless they are more massive than Jupiter. They speculate that there may even be a class of exoplanets that are the rocky cores of Jupiter-sized planets stripped of their atmospheres.

There is evidence of evaporative loss of volatiles on a planetary scale in our own solar system. Venus, often called Earth's twin due to their similar sizes, is much more depleted in water than the Earth. However, the D/H ratio of Venus' atmosphere suggests that it may have had water, which was evaporated away because of a higher surface temperature (Fegley 2004). Io, on the other hand, resides in the Jovian magnetosphere, an extreme radiation environment, and experiences extreme tidal heating. Io seems to have lost most of its H, C, and N, leaving S as the remaining dominant volatile element (Spencer & Schneider, 1996). The exoplanets considered here are in more extreme stellar environments than either Io or Venus and may experience more severe heating. We therefore do not think it unreasonable that some may have also lost their S contents. Detailed modeling of long-term volatile loss from terrestrial-type exoplanets should be explored in the future.

We focus here on modeling the planet CoRoT-7b. CoRoT-7b is the first known transiting super-Earth, making it an excellent candidate for potential atmospheric observations. Observations of CoRoT-7b ($M < 11 M_\oplus$, $a = 0.017$ AU) suggest that the planet is tidally-locked with a surface temperature of 1800 – 2600 K at the sub-stellar point on the day side (Léger et al. 2009). These temperatures are hot enough to melt and vaporize silicate material. In §2, we describe our computational methods. In §3, we describe the compositions and pressures of silicate atmospheres for different planetary compositions. In §4, we discuss some processes that may affect the silicate atmospheres and make recommendations for species that could be detected in the atmospheres of evaporating super-Earths. Preliminary results were given by Fegley & Schaefer (2005).

**2. COMPUTATIONAL METHODS**

We used the MAGMA code of Fegley & Cameron (1987) to model the composition of silicate atmospheres produced by vaporization of an Earth-like exoplanet. The MAGMA code, which is a robust and rapid mass-balance, mass action algorithm, is described and validated against experimental studies in Fegley & Cameron (1987) and Schaefer &

Fegley (2004a). The MAGMA code considers the elements Na, K, Fe, Mg, Si, Ti, Ca, Al, and O and their compounds. Computations are made for gas-melt equilibrium both with and without fractional vaporization (i.e., vapor is continuously removed). For CoRoT-7b, fractional vaporization simulates loss of material from the atmosphere. Loss may be due to hydrodynamic escape from the atmosphere or to transport of material from the day-side to the night-side, where condensation may effectively remove it from the system. We make no explicit assumptions about the mechanism of gas-loss, nor do we attempt to quantify the degree of loss.

We used temperatures of 1500 – 3000 K. The maximum temperature at the substellar point (1800 – 2600 K) of CoRoT-7b depends upon heat transport models (Léger et al. 2009), and the temperature drops significantly away from the sub-stellar point to ~150 K on the night side. The composition of CoRoT-7b is unknown, so we arbitrarily assume that the planet is a terrestrial-type planet rather than the core of a giant planet. We used a number of starting compositions, including: the Earth's continental and oceanic crusts, the bulk silicate Earth (BSE), and the bulk silicate Moon (Wedepohl 1995, Klein 2005, O'Neill & Palme 1998, Warren 2005). These compositions are given in Table 1 and are renormalized to 100% on a volatile-free basis.

## 3. RESULTS

Figure 1 shows the initial total atmospheric pressures (i.e., prior to any fractional vaporization) for our nominal silicate compositions as a function of temperature. Pressures decrease in the order: oceanic crust, BSE, continental crust, Moon. There is no clear correlation between elemental abundances and the initial total pressures, so the total pressure cannot be used to constrain the silicate composition. The total pressures decrease with temperature, from ~$10^{-1}$ bars at 3000 K to ~$10^{-7}$ bars at 1500 K. At 2200 K, the pressures range from ~$1\times10^{-3}$ to $5.5\times10^{-4}$ bars.

Figure 2 shows the composition of the atmosphere over the BSE as a function of temperature. Results are shown in terms of column density ($\sigma_i$, molecules/cm$^2$), which was calculated from the partial pressure of gas $i$ ($P_i$) using the equation:

$$\sigma_i = \frac{P_i N_A}{g\mu} \qquad (1)$$

where $N_A$ is Avogadro's number, $\mu$ is the molecular weight of species $i$, and $g$ is gravitational acceleration. We assumed the planetary properties for CoRot-7b ($R = 1.72 R_\oplus$, $M \sim 11 M_\oplus$) to calculate $g = 36.2$ m/s$^2$.

The major gas at all temperatures shown is monatomic Na, followed by $O_2$ and monatomic O. As temperature increases, the SiO abundance increases and approaches the abundances of the oxygen species. Magnesium and iron are present primarily as monatomic gases, and are less abundant than SiO gas. Singly ionized Na is produced by thermal ionization, and no contribution from photoionization is considered here. Column abundances are larger for the oceanic crust, but relative abundances are very similar to those for the BSE. For the continental crust and the Moon, the column abundance of SiO is larger relative to Na than in the BSE. The continental crust also has significant monatomic K, whereas monatomic Fe is more abundant in the atmosphere of the Moon. Column abundances for the Moon are significantly lower than for the other silicate compositions, as expected from the total pressures shown in Fig. 1.

Figure 3 shows how the atmospheric composition changes due to isothermal fractional vaporization. The initial total pressure in Fig. 1 and the gas chemistry in Fig. 2 are for 0% vaporization. 50% vaporization represents loss of 50% of the starting material, and so on. As gas is removed by fractional vaporization, the remaining silicate is enriched in more refractory elements (e.g., Ca, Al, and Ti). Figure 3a and 3b show the BSE and the continental crust, respectively, at 2200 K. Results for the Moon are similar to the BSE, and results for the oceanic crust are similar to the continental crust.

For the BSE, Na is completely lost from the system when only 2% has been vaporized. Potassium is the next to be lost at ~22% vaporization. After Na is lost, shown by the very rapid fall-off of Na at low fractions vaporized, SiO gas becomes the major species, followed by $O_2$, Fe and O gases. The mole fraction of SiO is greatest at ~16% vaporization and decreases slowly to ~90%, where it falls off rapidly. The oxygen species maintain a fairly constant abundance throughout. Monatomic Fe is most abundant at ~2%, then falls off but is not completely removed from the system until ~90% vaporization. Magnesium becomes the most abundant gas at ~27% vaporization and maintains a fairly constant abundance to ~90%. At greater than 90% vaporization, the gas is composed of Al, Ca, AlO, and $TiO_2$ gases. When Na is lost, the total pressure drops from ~$10^{-3}$ to ~$10^{-4}$ bars. Pressure is fairly constant from 2% to ~90% vaporization, then drops further to ~$10^{-6}$ to $10^{-7}$ bars.

At lower temperatures, elements are completely removed from the system at much lower vaporization fractions but persist to greater vaporization fractions at higher temperatures. For instance, Na is completely removed from the system at 1% and 11% vaporization at 1800 K and 2600 K, respectively, for the BSE. At 1800 K, once Na is lost, Fe gas is the most abundant vapor species until ~8% vaporization, when SiO gas becomes more abundant. Magnesium becomes the most abundant gas at ~22% vaporization. Results at higher temperatures are more similar to those shown in Fig. 3a.

Results for the continental and oceanic crusts are significantly different from the BSE, as shown in Fig. 3b. Sodium remains in the system to much higher fractions vaporized than in the BSE. Monatomic Na is the major gas until ~11% vaporization, beyond which SiO becomes more abundant, followed by $O_2$ and O. Sodium remains in the system to ~50% vaporization. The total pressure drops gradually, losing an order of magnitude by 63% vaporization. Potassium is a fairly abundant gas in the continental crust atmosphere at larger vaporization fractions, with a peak at ~67% and complete loss by ~74% vaporization. The oceanic crust contains less K, and Mg gas is less abundant than in the BSE. At lower temperatures in the continental crust, Na falls off more quickly, and Fe and K are less abundant in the gas and peak at lower vaporization fractions. At higher temperatures, SiO becomes the most abundant gas at smaller vaporization fractions, and Na persists to higher vaporization fractions. Additionally, Fe and K peak at higher vaporization fractions.

## 4. DISCUSSION

*4.1 Condensation and Clouds*

The atmospheric structure and composition of the exoplanets may be altered by silicate cloud condensation. The condensation clouds may eventually rain (or snow) out silicates onto the planet's surface. If condensates are reincorporated into the magma ocean, there is no net elemental fractionation. However, for a planet such as CoRoT-7b with a significant day-night difference, the condensates may be deposited on the night-side of the planet, effectively removing them from the magma ocean system. The general principle is similar to the metallic snow on Venus' mountain tops (Schaefer & Fegley 2004b)

We did simple condensation calculations for our BSE atmosphere at 2200 K, using a dry adiabatic lapse rate for a planet with $g = 36.2$ m/s$^2$. In these calculations, we assumed that the abundance of an element was reduced by formation of condensates. We did not consider the effect of condensation on the temperature/pressure profile (i.e., silicate wet adiabat). We considered condensation of metals, oxides, and compounds of no more than 2 oxides. Oxides and silicates of greater complexity are unlikely to form in a thin atmosphere due to kinetic inhibitions. These first order calculations indicate which elements are most susceptible to fractionation by vapor removal.

We found that Mg condensed primarily as enstatite $MgSiO_3$ (liq) at temperatures less than 2125 K ($z \sim 2$ km), with more than 90% condensed at altitudes below ~4 km. Aluminum formed both corundum ($Al_2O_3$, s) and spinel ($MgAl_2O_4$, s) and was 90% condensed below altitudes of ~6 km. Calcium, as wollastonite $CaSiO_3$ (s), was 90% condensed at altitudes less than ~7 km. Silicon condensed as $MgSiO_3$ (liq), $CaSiO_3$ (s), and silica $SiO_2$ (liq) below ~8 km. Iron condensed as "FeO" (liq) at altitudes below ~10 km. Titanium condensed as $CaTiO_3$ (s) and geikelite $MgTiO_3$ (liq) and was 90% condensed below ~15 km. Sodium and oxygen condense primarily as $Na_2O$ and are both completely condensed at an altitude of ~28 km. Potassium is the only element remaining in the gas at higher altitudes, and it finally condenses as metal, due to the lack of oxygen, at an altitude of ~50 km.

Given their low condensation altitudes, we believe that Mg, Al, Ca, Si, and Fe are likely to be primarily reincorporated into the magma ocean. Titanium and possibly Fe may be slightly depleted by removal to the night side of the planet. Sodium and potassium remain in the atmosphere to much higher altitudes, and their abundances may be more affected by photochemistry than by condensation.

*4.2 Extended Na and K Clouds*

Interaction of the stellar wind with the atmosphere may generate a large Na (and K) cloud around the planet. Clouds of Na have been observed in our solar system around Mercury and Io. At Mercury, the Na atmosphere (~$10^{11}$ cm$^{-2}$) is generated by impact vaporization and sputtering of the surface (Potter & Morgan, 1987) and extends to

~23$R_{Mercury}$ in the tail. On Io, volcanic eruptions emit Na and NaCl into the atmosphere, where it interacts with the Jovian magnetosphere (Spencer & Schneider 1996). The extended Na cloud of Io is a very bright feature, with an abundance of ~$10^{10}$-$10^{12}$ atoms Na cm$^{-2}$ near Io and extends out to ~500$R_{Jupiter}$ (~19,600$R_{Io}$) (Schneider et al. 1991). In comparison, the abundance of Na in our BSE atmosphere is $10^{15}$ cm$^{-2}$ at 1500 K up to $10^{20}$ cm$^{-2}$ at 3000 K. We therefore suggest that large neutral Na and possibly K clouds may be observable features around super-Earths in close proximity to their stars. A very large cloud would occult more of the star as the planet passes in front of it, which would make detection easier than for a closely bound atmosphere (Ehrenreich et al. 2006). Sodium has already been detected in the atmospheres of two giant exoplanets HD209458b and HD189733b (Charbonneau et al. 2002, Redfield et al. 2008), so we are hopeful that similar techniques may also observe Na in the atmospheres of the hot super-Earths.

However, the abundances of Na and K above the clouds may be significantly reduced by photoionization. CoRoT-7b orbits a G8V star, which is slightly cooler than the Sun but spectrally similar. Most exoplanets orbit main sequence stars of types F, G, K, and M, in order of decreasing luminosity. At such short orbital periods, however, the EUV flux should be much greater than that received by the Earth's atmosphere. Photoionization of Na I occurs at wavelengths shortward of 240 nm, and of K I at wavelengths shortward of 285 nm. The abundance of monatomic K is therefore likely to be more reduced than that of Na I. However, a detailed model of the photoionization of either element is outside the scope of this paper.

*4.3 Chemistry of Silicon Monoxide Gas*

Our results predict that SiO is the major Si-bearing gas in the atmospheres of hot super-Earths. Silicon monoxide has strong IR bands at 4 µm and 8 µm and lines throughout the millimeter region, and is the most widely observed Si-bearing molecule in space, with column densities of ~$10^{11}$-$10^{12}$ cm$^{-2}$ in diffuse and spiral-arm clouds (Lucas & Liszt 2000), and ~$10^{14}$-$10^{15}$ cm$^{-2}$ in molecular clouds, star-forming regions and cool stars (Groesbeck et al. 1994, Hatchell et al. 2001, Minh et al. 1992). In comparison, the calculated abundances for our BSE atmosphere are >$10^{15}$ cm$^{-2}$ at temperatures above 2000 K.

- 10 -

The destruction of SiO in the atmosphere may be initiated by absorption of UV light shortward of 300 nm (Podmoshenskii et al. 1968, Vorypaev 1981, Matveev 1986)

$$SiO + h\nu \rightarrow Si + O \quad \lambda < 300 \text{ nm} \qquad (2)$$

Photoionization of SiO occurs shortward of 107 nm (Baig & Connerade 1979) and is probably a much less important sink. Condensation of SiO as $MgSiO_3$ or $SiO_2$ is plausibly the most important loss reaction for SiO gas.

*4.4 Volatiles*

Using Io as an analogy for an evaporating terrestrial-type planet, the volatiles most likely to remain are S and Cl. Sulfur and Cl are the 10th and 19th most abundant elements in our solar system. Both elements remain sequestered in solid phases to higher temperatures than H, C, or N, making them less likely to be lost from the atmosphere. Additionally, S (32 g/mole) and Cl (35 g/mole) are significantly heavier than H, C, or N and would require much more energy to escape from the atmosphere. However, if any S or Cl remain, then the metal-bearing vapors produced by the evaporating magma ocean, primarily Na, $O_2$, and O, would be depleted by reactions with any S- or Cl-bearing species in the atmosphere. If we assume the S/Na (0.055) and Cl/Na (0.0075) ratios of the BSE, we find that the major Cl and S species are $SO_2$ and NaCl, respectively. Sulfur reacts with $O_2$ via the net thermochemical reaction:

$$\tfrac{1}{2} S_2(g) + O_2(g) = SO_2(g) \qquad (3)$$

At high temperatures, $SO_2$ is the third most abundant gas, after Na and $O_2$. Chlorine reacts with Na via the net thermochemical reaction:

$$Na(g) + \tfrac{1}{2} Cl_2(g) = NaCl(g) \qquad (4)$$

At 2200 K, NaCl is the fifth most abundant gas, after Na, $O_2$, $SO_2$, and O. As temperature drops, NaCl becomes more abundant than monatomic O. The abundances of Cl and S assumed here are low enough that they do not significantly reduce the column abundances of Na or $O_2$ at high temperatures. If the Cl/Na or S/Na ratios are significantly larger than our assumed values, the abundances of Na and $O_2$ may be greatly reduced.

Sulfur and Cl may also alter the condensation sequence of the clouds. In the presence of S, Na may condense as $Na_2SO_4$, but likely not as $Na_2S$. Clouds of $Na_2S$ are predicted to exist on HD209458b (Charbonneau et al. 2002), and have long been predicted in the atmospheres of Jupiter and brown dwarfs (Fegley & Lodders 1994,

Lodders 1999). These gas giants and brown dwarfs retain their hydrogen; therefore, their atmospheres are significantly more reducing than that of a hot super-Earth. Sodium sulfate condenses at higher temperatures and lower altitudes than $Na_2O$ and would deplete the abundance of Na in the upper atmosphere. In the presence of Cl, K may condense as KCl at altitudes of ~35 km, compared to 50 km for K metal.

*4.5 Recommended Species to Search for in the Atmospheres of Hot Super-Earths*

We conclude that the major gases in a hot super-Earth stripped of its volatiles will be monatomic Na gas, $O_2$, O, and SiO gas. The abundances of these gases may be depleted by condensation, photoionization and reaction with any volatiles remaining in the atmosphere. Sodium may form a large cloud around the exoplanet, similar to that seen around Io. Potassium, although not a major gas, may be an important constituent in the upper atmosphere due to its low condensation temperature. Models of long-term loss of volatiles from an evaporating terrestrial-type exoplanet could help further shape predictions for the atmospheric compositions of newly discovered hot super-Earths.

## 5. ACKNOWLEDGMENTS

This work is supported by Grant NNG04G157A from the NASA Astrobiology Institute.

# 6. REFERENCES

Baig, M. A. & Connerade, J. P. 1979, J. Physics B, 12, 2309.

Baraffe, I., Chabrier, G., Barman, T. S., Selsis, F., Allard, F., & Hauschildt, P. H. 2005, A&A, 436, L47.

Baraffe, I., Selsis, F., Chabrier, G., Barman, T. S., Allard, F., Hauschildt, P. H., & Lammer, H. 2004, A&A, 419, L13.

Barnes, R., Jackson, B., Raymond, S. N., West, A. A., & Greenberg, R. 2009, ApJ, 695, 1006.

Charbonneau, D., Brown, T. M., Noyes, R. W., & Gilliland, R. L. 2002, ApJ, 568, 377.

Ehrenreich, D., Tinetti, G., Lecavelier des Etangs, A., Vidal-Madjar, A. & Selsis, F. 2006, A&A, 448, 379.

Elkins-Tanton, L. T., & Seager, S. 2008, ApJ, 685, 1237.

Fegley, B., Jr. 2004, in Meteorites, Comets, and Planets, ed. A. M. Davis, Vol. 1 Treatise on Geochemistry, ed. K.K. Turekian & H. D. Holland (Boston, MA: Elsevier) 487.

Fegley, B., Jr., & Lodders, K. 1994, Icarus, 110, 117.

Fegley, B., Jr. & Schaefer, L. 2005, BAAS, 37, 683.

Groesbeck, T. D., Phillips, T. G. & Blake, G. A. 1994, ApJ, 94, 147.

Ganesan, A. L., Elkins-Tanton, L. T., & Seager, S. 2008, LPSC, 39, abstract 1368.

Hatchell, J., Fuller, G. A., & Millar, T. J. 2001, A&A, 372, 281.

Hubbard, W. B., Hattori, M. F., Burrows, A., Hubeny, I., & Sudarsky, D. 2007, Icarus, 187, 358.

Jackson, B., Barnes, R., & Greenberg, R. 2008a, Mon. Not. R. Astron. Soc., 391, 237.

Jackson, B., Greenberg, R., & Barnes, R. 2008b, ApJ, 681, 1631.

Klein, E. M. 2005 in The Crust, ed. R. Rudnick, Vol. 3. Treatise on Geochemistry, ed. K.K. Turekian & H.D. Holland (Boston, MA: Elsevier) 433.

Lecavelier des Etangs, A. 2007, A&A, 461, 1185.

Lecavelier des Etangs, A., Vidal-Madjar, A., McConnell, J. C., & Hébrard, G. 2004, A&A, 418, L1.

Léger, A. et al. 2009, A&A, submitted.

Lodders, K. 1999, ApJ, 519, 793.

Lucas, R., & Liszt, H. S. 2000, A&A, 355, 327.

Matveev, V. S. 1986, J. Appl. Spectros., 45, 183.

Miller-Ricci, E., Seager, S., & Sasselov, D. 2009, ApJ, 690, 1056.

Minh, Y. C., Irvine, W. M., & Friberg, P. 1992, A&A, 258, 489.

O'Neill, H. St. C. & Palme, H. 1998, in The Earth's Mantle-Composition, Structure and Evolution, ed. I. Jackson (Cambridge, UK: Cambridge University Press) 3.

Podmoshenskii, I. V., Polozova, L. P., Chirkov, V. N., & Yakovleva, A. V. 1968, J. Applied Spectros., 9, 707.

Potter, A. E. & Morgan, T. H. 1987, Icarus, 71, 472.

Redfield, S., Endl, M., Cochran, W. D., & Koesterke, L. 2008, ApJ, 673, L87.

Schaefer, L., & Fegley, Jr., B. 2004a, Icarus, 169, 216.

Schaefer, L., & Fegley, Jr., B. 2004b, Icarus, 168, 215.

Schneider, N. M., Hunten, D. M., Wells, W. K., Schultz, A. B. & Fink, U. 1991, ApJ, 368, 298.

Spencer, J. R. & Schneider, N. M. 1996, Ann. Rev. Earth Planet. Sci., 24, 125.

Vidal-Madjar, A., Lecavelier des Etangs, A., Désert, J.-M., Ballester, G. E., Ferlet, R., Hébrard, G., & Mayor, M. 2003, Nature, 422, 143.

Vidal-Madjar, A. et al. 2004, ApJ, 604, L69.

Vorypaev, G. G. 1981, J. Applied Spectros., 35, 1297.

Warren, P. H. 2005, M&PS, 40, 477.

Wedepohl, K. H. 1995, GCA, 59, 1217.

Table 1. Compositions of planetary analogs.

| Oxide (wt%) | continental crust[a] | oceanic crust[b] | BSE[c] | Moon[d] |
|---|---|---|---|---|
| $SiO_2$ | 62.93 | 50.36 | 45.97 | 47.17 |
| MgO | 3.79 | 7.61 | 36.66 | 36.29 |
| $Al_2O_3$ | 15.45 | 15.85 | 4.77 | 3.90 |
| $TiO_2$ | 0.70 | 1.48 | 0.18 | 0.18 |
| FeO | 5.78 | 9.56 | 8.24 | 9.31 |
| CaO | 5.63 | 12.24 | 3.78 | 3.08 |
| $Na_2O$ | 3.27 | 2.77 | 0.35 | 0.05 |
| $K_2O$ | 2.45 | 0.13 | 0.04 | 0.005 |
| Total | 100.00 | 100.00 | 99.99 | 99.985 |

[a]Wedepohl (1995). [b] Klein (2005). [c]BSE = Bulk Silicate Earth. O'Neill & Palme (1998). [d]Warren (2005).

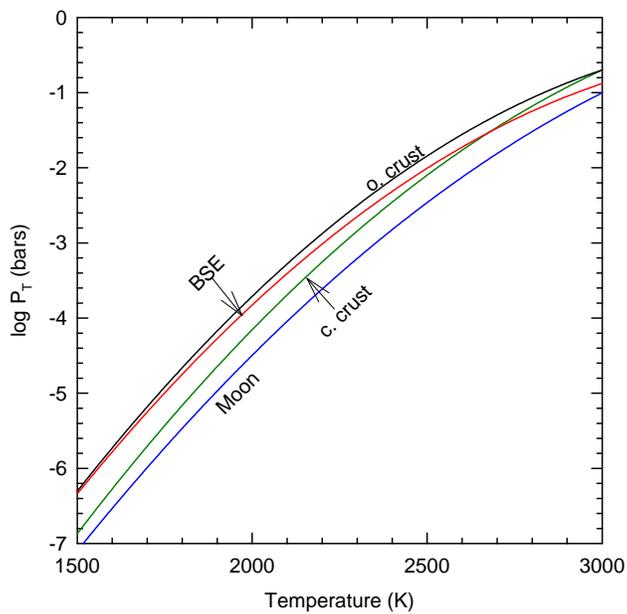

**Figure 1.** Initial atmospheric pressure as a function of temperature for planetary analogs (c. crust = continental crust, o. crust = oceanic crust, BSE = bulk silicate Earth, Moon = bulk silicate Moon, see Table 1).

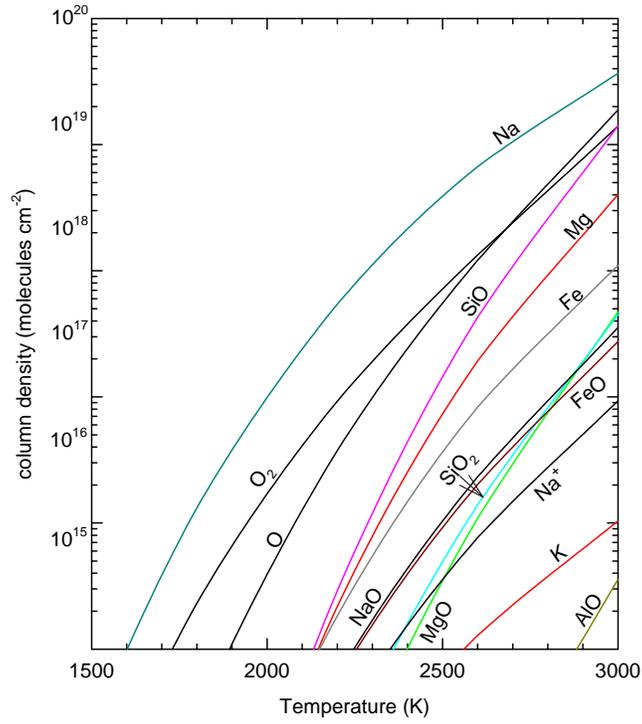

**Figure 2.** Atmospheric composition as a function of temperature for the bulk silicate Earth (BSE) as column densities (molecules/cm$^2$) for a planet with $g = 36.2$ m/s$^2$.

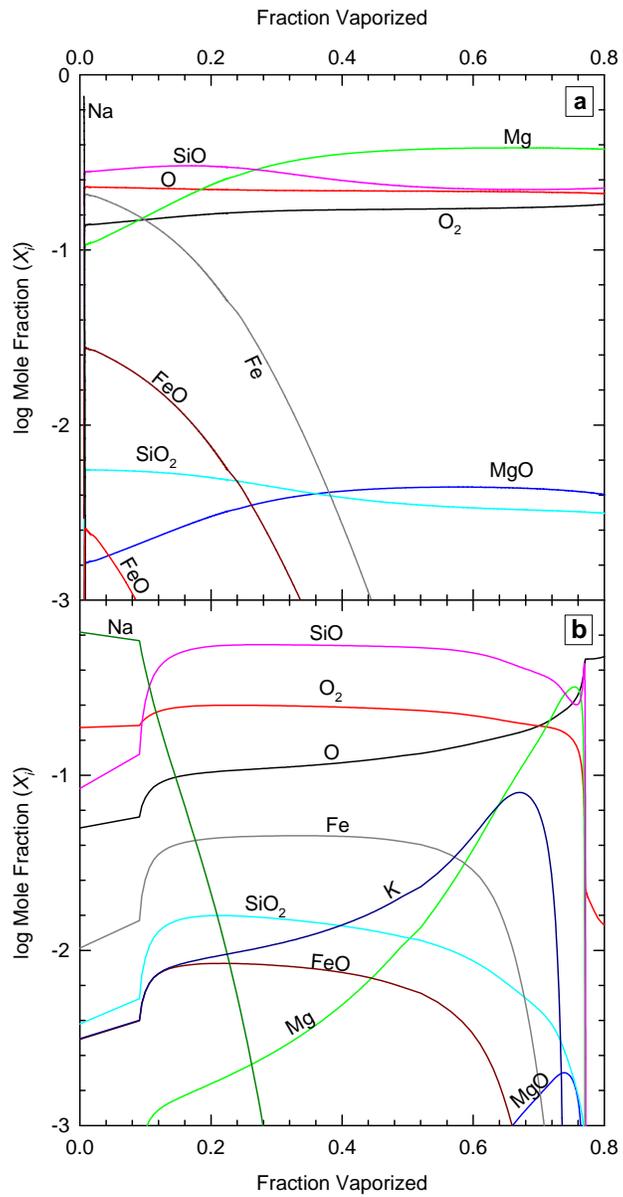

**Figure 3.** Composition of the atmosphere of (a) the bulk silicate Earth (BSE) and (b) the continental crust in mole fractions during fractional vaporization at 2200 K. The fraction vaporized is given by mole.